\documentclass[a4paper,12pt]{article} 
\usepackage[T1]{fontenc}
\usepackage[utf8]{inputenc} 
\usepackage{amsmath} 
\usepackage{amsfonts}
\usepackage{amssymb} 
\usepackage[english]{babel}
\usepackage{graphicx,color}
\begin{document}
\title{Vacuum fluctuations in the presence of nonlinear boundary 
conditions} 
\author{C.~D.~Fosco$^{a}$ and L.~E.~Oxman$^{b}$\\
{\normalsize\it $^a$Centro At\'omico Bariloche and Instituto Balseiro}\\
{\normalsize\it Comisi\'on Nacional de Energ\'\i a At\'omica}\\
{\normalsize\it R8402AGP Bariloche, Argentina.}\\ 
{\normalsize\it $^b$ Instituto de F\'{\i}sica}\\ 
{\normalsize\it Universidade Federal Fluminense}\\ {\normalsize\it Campus
da Praia Vermelha}\\ 
{\normalsize\it Niter\'oi, 24210-340, RJ, Brazil.}} 
\date{} 
\maketitle 
\begin{abstract} 
We consider a system consisting of a quantum, massless, real scalar field,
in the presence of {\em nonlinear\/} mirrors: infinite parallel planes,
upon which the field satisfies nonlinear boundary conditions. The latter
are implemented by non-quadratic interaction vertices, strictly localized
on the mirrors.  By using the appropriate perturbative expansions, we
obtain approximate expressions for the Casimir energy corresponding to weak
coupling, regarding the strength of the interaction terms. We also comment
on an alternative expansion scheme that may be useful when the weak coupling
expansion is not justified.
\end{abstract}
\section{Introduction}\label{sec:intro} 
Quantum vacuum fluctuations may manifest themselves, under the proper
circumstances, in the form of observable macroscopic physical effects.  For
instance, the existence of boundary conditions for a quantum field on a
nontrivial geometry can produce a Casimir force, an effect which has been
evaluated for different kinds of vacuum fields, geometries, and boundary
conditions~\cite{rev}.  Among the many important developments that have
emerged in those areas, a topic which has recently received much attention
has been the use of a more accurate description of the `mirrors'; in other
words, of the geometry and nature of the boundary conditions. 

A refined description may include corrections that are an attempt to
represent, for example, a departure from the idealized situation of exactly
conducting, zero-width, smooth-shape mirrors. Examples of those corrections are:
roughness, finite temperature and conductivity, as well as a finite width.

Leaving aside the case of nonlocal boundary
interactions~\cite{Elizalde:1995bb,Saharian:2006zm,Fosco:2009cw}, for which
extensive studies have been carried
out~\cite{EsquivelSirvent:2006iw,bajnok}, imperfect boundary conditions are
usually represented, at least in some scalar field models, by the
introduction of a `space dependent mass term', such that the -otherwise
massless- scalar field becomes massive just at the locii of the
mirrors~\cite{Graham:2003ib}, therefore `penalizing' the development of
non-zero field values  on the mirror. Dirichlet boundary conditions appear,
in this context, wherever that space-dependent mass tends to
infinity~\cite{Jaffe:2003ji}.

In this kind of model, the relevant properties of the mirrors correspond to  a
linear response approximation.  A `microscopic' way to interpret this
approximation, in quantum field theory (QFT) terms, amounts to using a
truncated version of the expansion of the effective action, for the vacuum
field, due to the microscopic degrees of freedom living on the mirrors, to
the second order in the vacuum field~\cite{Fosco:2008td,Fosco:2009zc}.

On the other hand, neglecting higher-order terms in the expansion can be
expected to be a reliable approximation whenever the magnitude of the
quadratic term is large in comparison with the subleading, usually quartic,
one. Indeed, since a large quadratic term penalizes the existence of a non
zero field configuration around the corresponding mirror, the contributions
from higher-order terms (since they involve higher powers of the field) are
likely to be suppressed.  
It is the main purpose of this paper to study the consequences for the Casimir
energy of the presence of that kind of nonlinearity, having in mind cases
where the conditions to discard it are not necessarily met. 

This paper is organized as follows: in Section~\ref{sec:themodel}, we
define the model and its relevant properties, and in
Section~\ref{sec:vacuum} we consider its vacuum energy within the context
of the functional integral formalism. Then, in Section~\ref{sec:weak}, we
present a study of the weak coupling expansion for the nonlinearities.
In section~\ref{sec:mixed} we consider a weak coupling expansion adapted to
the case when boundary conditions have both linear and nonlinear
parts. An alternative, complementary expansion is introduced and considered
in Section~\ref{sec:strong}; it corresponds to a situation where there is a
small correction on top of a Dirichlet boundary condition.  
Finally, in Section~\ref{sec:conclusions}, we present our conclusions.

\section{The model}\label{sec:themodel} 
The model that we consider involves a real massless scalar vacuum field
$\varphi(x)$ in $3+1$ dimensions, coupled to two flat, parallel, zero-width
mirrors, denoted by $L$ and $R$, which occupy the planes $x_3 = 0$ and
$x_3=a$, respectively. Euclidean coordinates shall be denoted by $x \equiv
(x_0,x_1,x_2,x_3)$, and we will also use an specific notation, $x_\parallel
\equiv (x_0,x_1,x_2)$, for the coordinates which are parallel to the
mirrors' planes (on which we assume the existence of translation
invariance).  Alongside the last convention, we shall use letters from the
beginning of the Greek alphabet ($\alpha, \beta,\ldots$) to denote indices
which run over the values $0$, $1$, and $2$.

The media inside each mirror is assumed to be strictly confined to the
respective plane, so that the full Euclidean action for the system,
${\mathcal S}(\varphi)$, naturally decomposes as follows:
\begin{equation}\label{eq:defs} 
{\mathcal S}(\varphi)\;=\; {\mathcal S}_f(\varphi) \,+\,{\mathcal
S}_I(\varphi) \;, 
\end{equation} 
where
${\mathcal S}_f$ denotes the free action (i.e, in the absence of mirrors)
for the real scalar field: 
\begin{equation}\label{eq:defsf} 
{\mathcal S}_f(\varphi) \;=\; \frac{1}{2} \int_x \big(\partial \varphi
\big)^2 \;,
\end{equation} 
and 
\begin{equation}\label{eq:defsi} 
{\mathcal S}_I(\varphi)\;=\;{\mathcal S}_L(\varphi) \,+\,{\mathcal
S}_R(\varphi) \;,
\end{equation} 
while ${\mathcal S}_L$ and ${\mathcal S}_R$ account for the coupling
between $\varphi$ and the respective mirror. These terms shall be assumed
to have a similar structure. We will endow them with, for the sake of
simplicity, a {\em local\/} form, confined to a $2+1$-dimensional
spacetime, the world-volumes generated by the static mirrors during the
course of (trivial) time evolution: 
\begin{equation}\label{eq:defslsr}
{\mathcal S}_L \;=\; \int_{x_\parallel}\, {\mathcal
V}_L[\varphi(x_\parallel,0)] \;\;,\;\;\;\; {\mathcal S}_R \;=\;
\int_{x_\parallel} \,{\mathcal V}_R[\varphi(x_\parallel,a)] \;, 
\end{equation} 
where ${\mathcal V}_L$ and ${\mathcal V}_R$ are local
functions of their arguments, involving no derivatives of the fields.

From the classical equations of motion that follow from the {\em real-time\/}
version of the action for $\varphi$, we see that they imply the boundary
conditions: 
\begin{equation} 
\left\{ \begin{array}{ccc}
\partial_3\varphi(x_\parallel,0_+) - \partial_3\varphi(x_\parallel,0_-)
&=& \frac{\partial {\mathcal V}_L}{\partial
\varphi}\big[\varphi(x_\parallel,0)\big] \\ 
\partial_3\varphi(x_\parallel,a_+) - \partial_3\varphi(x_\parallel,a_-) 
&=& \frac{\partial {\mathcal V}_R}{\partial
\varphi}\big[\varphi(x_\parallel,a)\big] \;,  
\end{array} \right.
\end{equation}
which necessarily introduce nonlinearities as soon as one of the functions
${\mathcal V}_L$ or  ${\mathcal V}_R$ involves more than two powers of its
argument.
A consequence of this very same property is the following: when there are
more than two powers of the field in one of the mirrors, the quantum
equations of motion for $\varphi$, namely, the equations for its mean
value  $\langle\varphi\rangle$, are different to their classical
counterparts. 
Indeed, from: $0 = \int {\mathcal D}\varphi
\,\frac{\delta}{\delta\varphi(x)} e^{-{\mathcal S}(\varphi)}$,
we obtain
\begin{equation}
\Box \langle \varphi(x)\rangle + \delta(x_3) \big\langle \frac{\partial
{\mathcal V}_L}{\partial \varphi}[\varphi(x_\parallel,0)]\big\rangle +
\delta(x_3 - a) \big\langle \frac{\partial {\mathcal V}_R}{\partial
\varphi}[\varphi(x_\parallel,a)]\big\rangle\;,
\end{equation}
from which linear boundary conditions are obtained only when the 
potentials ${\mathcal V}_{L,R}$ are quadratic. On the contrary, for
non-quadratic interactions the equation involves Green's function with more
than one field, and the resulting system of equations does not close; in
other words, it becomes infinite.

\section{Vacuum energy}\label{sec:vacuum} 
The vacuum energy $E$ will be obtained from the effective action $\Gamma$
(for the static configuration of two parallel planes already defined) when
evaluated for a long, yet finite, time interval of length $T$: 
\begin{equation} 
E \;=\; \lim_{T \to \infty} \;\frac{\Gamma}{T} \;, 
\end{equation} 
where $\Gamma = - \log {\mathcal Z}$, and ${\mathcal Z}$ denotes the
Euclidean vacuum transition amplitude: 
\begin{equation}\label{eq:defz} 
{\mathcal Z}\;=\; \int {\mathcal D}\varphi \,
e^{- {\mathcal S}(\varphi)} \;, 
\end{equation} 
and the action is evaluated on the (Euclidean) time interval
\mbox{$[-\frac{T}{2},\frac{T}{2}]$}.

By factoring out the partition function corresponding to ${\mathcal S}_f$,
we see that ${\mathcal Z}$ may be rewritten in the equivalent way:
\begin{equation}\label{eq:exp1} 
{\mathcal Z} \;=\;{\mathcal Z}_f \times
{\mathcal Z}_I 
\end{equation} 
with 
\begin{equation} 
{\mathcal Z}_f\;=\;
\int {\mathcal D}\varphi \, e^{- {\mathcal S}_f(\varphi)} \;,
\end{equation} 
and 
\begin{equation} 
{\mathcal Z}_I\;=\; \langle e^{- {\mathcal S}_I(\varphi)} \rangle 
\end{equation} 
where we have introduced a `Gaussian average' symbol $\langle \ldots
\rangle$ to denote the functional averaging with the weight defined by
${\mathcal S}_f$, namely, for any expression, its average is given by: 
\begin{equation}\label{eq:defav}
\langle \ldots \rangle \;\equiv\; \frac{\int {\mathcal D}\varphi \ldots
e^{- {\mathcal S}_f(\varphi)}}{\int {\mathcal D}\varphi \, e^{- {\mathcal
S}_f(\varphi)}} \;.  
\end{equation} 
Then, 
\begin{equation} 
\Gamma \;=\; \Gamma_f \,+\, \Gamma_I 
\end{equation} 
where \mbox{$\Gamma_f = -\log {\mathcal Z}_f$} and \mbox{$\Gamma_I = -\log
{\mathcal Z}_I$}.  Note that $\Gamma_f$ yields the vacuum energy
corresponding to the free field system, while $\Gamma_I$ contains
contributions due to the presence of the boundary conditions.

Since the Casimir force is insensitive to $\Gamma_f$, we shall discard that
contribution in what follows.  Besides, note that $\Gamma$, $\Gamma_f$ and
$\Gamma_I$ are not only proportional to $T$, but also to $L^2$, the area of
the mirrors.  This is just a manifestation of the fact that the system has
translation invariance along the two parallel directions to the mirrors, as
well as being time independent.
Thus rather than working with energies, which are extensive and therefore
proportional to $L^2$, we will use instead energies per unit area, denoted by
${\mathcal E}$, ${\mathcal E}_f$, and ${\mathcal E}_I$, respectively.
Thus, the interesting quantity shall be: 
\begin{equation} 
{\mathcal E}_I \;=\; - \lim_{T,L \to \infty}  \left[ \frac{1}{T L^2} \, \log \langle e^{-
{\mathcal S}_I(\varphi)} \rangle \right]\;.  
\end{equation}

Since we just need to keep terms that do contribute to the Casimir force,
we can also subtract from $E_I$ contributions which, although sensitive to
the existence of the boundary conditions, are independent of the distance
$a$ between the mirrors. That is the case of the mirrors' self-energies
which, although certainly may depend on the details of each interaction
term ${\mathcal S}_L$ and ${\mathcal S}_R$, are independent of the distance
$a$ between $L$ and $R$.

\section{Weak coupling expansion}\label{sec:weak} 
Let us calculate here the contribution to the Casimir energy due to purely
nonlinear coupling terms, under the assumption that those terms are small.  
The procedure is entirely analogous, although applied to a nonlinear
medium, to the approach followed, for example, in~\cite{Milton:2008vr} and 
\cite{Wagner:2008qq} to derive exact results in the weak coupling regime of
the static Casimir effect.

The perturbative expansion, taking as zeroth order the free
action ${\mathcal S}_f$, amounts to expand $\Gamma_I$ in powers of
${\mathcal S}_I$, 
\begin{equation} 
\Gamma_I \;=\; \Gamma_I^{(1)} \,+\,\Gamma_I^{(2)} \,+\,\ldots \;, 
\end{equation} 
where the superscript denotes the order (in ${\mathcal S}_I$) of each term
in the perturbative expansion.

Up to the second order, the explicit form of the terms is as follows:
\begin{equation} 
\Gamma_I^{(1)}\,=\, \langle {\mathcal S}_I \rangle \;,
\end{equation} 
and 
\begin{equation}
\Gamma_I^{(2)}\,=\, - \frac{1}{2} \Big[ \langle ({\mathcal S}_I)^2 \rangle
\,-\, \langle {\mathcal S}_I \rangle^2 \Big] \;.  
\end{equation}

Regarding the first-order term, we note that, since the quantum averaging
is defined with the free action, the result is a sum of two self-energy
terms, each one independent of the distance $a$ between the two mirrors. Thus,
there is no contribution from this term to the Casimir interaction energy. 

Let us now consider the second order term for the concrete example of
mirrors described by the terms: 
\begin{equation}\label{eq:defnlint}
{\mathcal S}_L \;=\; \int_{x_\parallel} \, \frac{g_L}{k_L!} \,
:[{\varphi}(x_\parallel,0) ]^{k_L}: \;\;,\;\;\; {\mathcal S}_R \;=\;
\int_{x_\parallel} \, \frac{g_R}{k_R!} \,
:[{\varphi}(x_\parallel,a)]^{k_R}:
\;,  
\end{equation} 
where the normal-order symbol means that contractions at the same vertex
are to be discarded~\footnote{The normal order symbol could be dropped at
the expense of adding a certain number of counterterms, which form a polynomial of
degree $k_{L,R}-2$ at the respective mirror, having the parity of the
integer $k_{L,R}$.}. 

The second-order term can then be written as follows: 
\begin{equation} 
\Gamma_I^{(2)}\,=\, -  \langle {\mathcal S}_L \; {\mathcal S}_R  \rangle \;. 
\end{equation}
More explicitly, one can see that $\Gamma_I^{(2)}$ vanishes unless $k_L = k_R \equiv k$, and 
\begin{eqnarray}
\Gamma_I^{(2)} &=& - \frac{g_L g_R}{k!} \, \int_{x_\parallel, x'_\parallel}
[ \langle {\varphi}(x_\parallel,a) {\varphi}(x'_\parallel,0) \rangle ]^k
\nonumber\\ &=& - \frac{g_L g_R}{k!} \, T L ^2 \, \int_{x_\parallel} [
\langle {\varphi}(x_\parallel,a) {\varphi}(0_\parallel,0) \rangle ]^k \;,
\end{eqnarray} 
where 
\begin{equation}
\langle {\varphi}(x_\parallel,x_3) {\varphi}(y_\parallel,y_3) \rangle \;=\;
\frac{1}{4 \pi^2 \; [(x_\parallel-y_\parallel)^2 + (x_3-y_3)^2]} \;.
\end{equation}
The fact that the non vanishing contributions to the Casimir interaction
appear only for $k_L = k_R$ is represented in Figure 1, for the particular case $k_L = k_R = 4$.

A rather straightforward calculation yields for the
interaction energy per unit area: 
\begin{equation} 
{\mathcal E}_I^{(2)}
\;=\; - \frac{g_L g_R}{k!} \, \frac{\pi^{1/2} \Gamma( 2 k - 3/2)}{ 2
(2\pi)^{2k-1} \Gamma(k)} \; \frac{1}{a^{2 k - 3}} \;, 
\end{equation}
which is an expression formally valid for any $k > 3/4$.
\begin{figure}[h!]
\begin{center}
\begin{picture}(0,0)%
\includegraphics{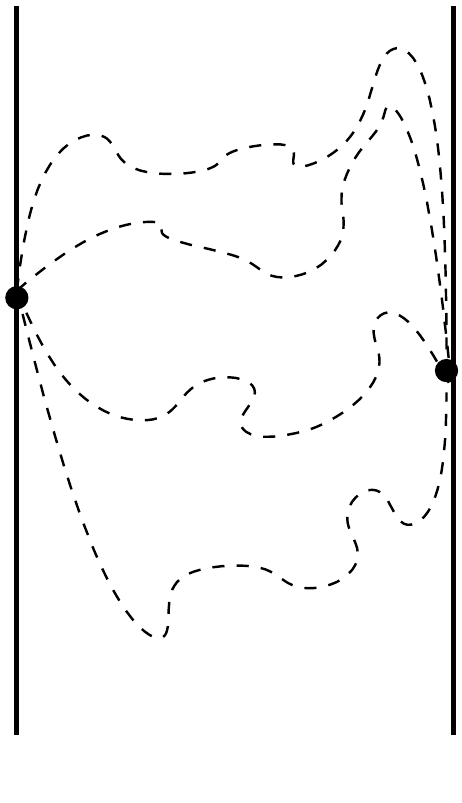}%
\end{picture}%
\setlength{\unitlength}{3066sp}%
\begingroup\makeatletter\ifx\SetFigFont\undefined%
\gdef\SetFigFont#1#2#3#4#5{%
  \reset@font\fontsize{#1}{#2pt}%
  \fontfamily{#3}\fontseries{#4}\fontshape{#5}%
  \selectfont}%
\fi\endgroup%
\begin{picture}(2838,4948)(796,-4976)
\put(811,-4921){\makebox(0,0)[lb]{\smash{{\SetFigFont{12}{14.4}{\rmdefault}{\mddefault}{\updefault}{\color[rgb]{0,0,0}$L$}%
}}}}
\put(3556,-4921){\makebox(0,0)[lb]{\smash{{\SetFigFont{12}{14.4}{\rmdefault}{\mddefault}{\updefault}{\color[rgb]{0,0,0}$R$}%
}}}}
\end{picture}%
\end{center}
\label{figure1}
\caption{The first nontrivial correction in the weak coupling regime, for
$k=4$.}
\end{figure}
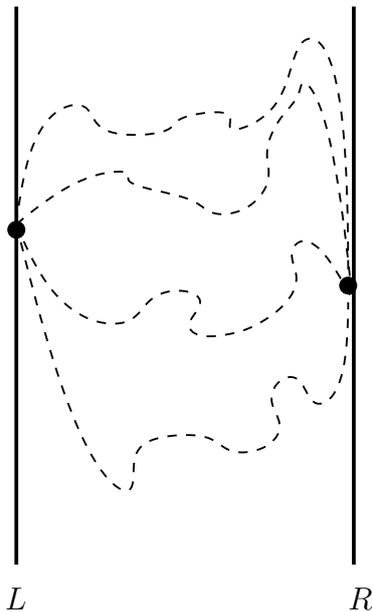
It is interesting to note that, for `perfect' boundary conditions
(Dirichlet, for instance), the only dimensionful quantity which appears in
the energy density is the distance $a$; this implies that ${\mathcal E}_I
\propto \frac{1}{a^3}$. 
On the other hand,  the perturbative result above should be valid for weak
coupling, and one cannot therefore take the limit when the couplings tend to
infinity.
However, we may note
that, since the mass dimension of the coupling constants is $[M]^{3-K}$,
when $k=3$ they become dimensionless. Thus, in that situation
the dependence with $a$ is entirely analogous to the case of Dirichlet or
Neumann boundary conditions, since:
\begin{equation} 
\big[{\mathcal E}_I^{(2)}\big]|_{k->3} \;=\; - \frac{g_L g_R}{1536 \pi^4\, a^3} \;. 
\end{equation}
It is rather straightforward to check that, for the $k=3$ case, the third
order term vanishes, except for self-energy contributions.

The weak coupling approximation can also be applied to situations where the
interaction terms are not necessarily of a polynomial form. Indeed,
assuming that the functions ${\mathcal V}_{L,R}$ can be represented in
terms of their Fourier transforms, ${\widetilde{\mathcal V}}_{L,R}$,
respectively:
\begin{eqnarray}
{\mathcal V}_L[\varphi(x_\parallel,0)] &=& \int
\frac{d\lambda_L}{2\pi} \, {\widetilde{\mathcal V}}_L(\lambda_L) e^{i
\lambda_L \, \varphi(x_\parallel,0)} \nonumber\\
{\mathcal V}_R[\varphi(x_\parallel,a)] &=& \int \frac{d\lambda_R}{2\pi} \,
{\widetilde{\mathcal V}}_R(\lambda_R) e^{i \lambda_R \,
\varphi(x_\parallel,a)} \;,
\end{eqnarray} 
the normal-ordered versions of these objects are given by~\cite{ZinnJustin:2002ru}:
\begin{eqnarray}
:{\mathcal V}_L[\varphi(x_\parallel,0)]: &=& \int
\frac{d\lambda_L}{2\pi} \, {\widetilde{\mathcal V}}_L(\lambda_L) 
\; 
:e^{i \lambda_L \, \varphi(x_\parallel,0)}:
 \nonumber\\
:{\mathcal V}_R[\varphi(x_\parallel,a)]: &=& \int
\frac{d\lambda_R}{2\pi} \, {\widetilde{\mathcal V}}_R(\lambda_L) 
\;
:e^{i \lambda_R \, \varphi(x_\parallel,a)}: 
\end{eqnarray} 
where
\begin{equation}
:e^{i \lambda_L \, \varphi(x_\parallel,0)}: \;=\; e^{i \lambda_L \, \varphi(x_\parallel,0) + \frac{1}{2} \lambda_L^2 \langle
\varphi(x_\parallel,0)\varphi(x_\parallel,0)\rangle}
\end{equation}
and
\begin{equation}
:e^{i \lambda_R \, \varphi(x_\parallel,a)}: \;=\; e^{i \lambda_R \, \varphi(x_\parallel,a) + \frac{1}{2} \lambda_R^2 \langle
\varphi(x_\parallel,a)\varphi(x_\parallel,a)\rangle} \;.
\end{equation}

Thus, the first non-trivial contribution to the interaction energy comes
from:
\begin{eqnarray}
\Gamma_I^{(2)}&=& -\int\frac{d\lambda_L}{2\pi} \,\int
\frac{d\lambda_R}{2\pi} \, {\widetilde{\mathcal V}}_L(\lambda_L) 
{\widetilde{\mathcal V}}_R(\lambda_R) \;
\int_{x_\parallel,x'_\parallel}\, 
\langle 
:e^{i \lambda_L \, \varphi(x_\parallel,0)}:
\;
:e^{i \lambda_R \, \varphi(x'_\parallel,a) }:
\rangle \nonumber\\
&=& - T L^2 \int\frac{d\lambda_L}{2\pi} \,\int
\frac{d\lambda_R}{2\pi} \, {\widetilde{\mathcal V}}_L(\lambda_L) 
{\widetilde{\mathcal V}}_R(\lambda_R) \;\int_{x_\parallel} 
e^{- \lambda_L \lambda_R \, \langle
\varphi(x_\parallel,a) \varphi(0,0)\rangle}\;,
\end{eqnarray}
and the surface density of (interaction) energy may be put in the 
form:
\begin{equation}
{\mathcal E}_I^{(2)}\;=\; - 4 \pi \,
\int\frac{d\lambda_L}{2\pi}
\,\int \frac{d\lambda_R}{2\pi} \, {\widetilde{\mathcal V}}_L(\lambda_L) 
{\widetilde{\mathcal V}}_R(\lambda_R) \;\int_0^\infty dr r^2 \,
e^{- \frac{\lambda_L \lambda_R}{4\pi^2} \, 
\frac{1}{r^2 + a^2}}\;.
\end{equation}

It is interesting to see that the previous expression can be expanded for
large $a$, obtaining:
\begin{equation}
{\mathcal E}_I^{(2)}\;\sim\; 
\sum_{l=2}^\infty \frac{c_l}{a^{2l-3}}
\end{equation}
with:
\begin{equation}
c_l \;=\; \frac{(-1)^{l+1} \pi^{1/2}
\Gamma(l-3/2)}{2^{2l} \pi^{2l-1}\Gamma(l) l!} 
\big(\int\frac{d\lambda_L}{2\pi} \, {\widetilde{\mathcal V}}_L(\lambda_L)
\lambda_L^l \big)
\big(\int\frac{d\lambda_R}{2\pi} \, {\widetilde{\mathcal V}}_R(\lambda_R)
\lambda_L^l\big) \;,
\end{equation}
which depends on the momenta of the Fourier transform of the `potentials'.
This shows, in particular, that the wider the Fourier transform for the
potentials, the larger shall the (negative) degree in the distance $a$ of the terms
contributing to the energy.

\section{Linear plus nonlinear boundary conditions}\label{sec:mixed}
In this case, we consider mirrors which are described by coupling terms
including both quadratic and non-quadratic pieces. The former can and will be
treated here exactly, while the latter will be expanded in a perturbative
expansion.
For the quadratic part, we include mass-like terms for the fields at
the mirrors, and, in order to evaluate ${\mathcal Z}$ (and $\Gamma$), we shall use a perturbative
expansion in the nonlinear terms;  we split the full action into its
quadratic and quartic parts:
\begin{equation}\label{eq:defs0i}
{\mathcal S}(\varphi) \;=\; {\mathcal S}_0(\varphi) \,+\, {\mathcal
S}_I(\varphi) \;,
\end{equation}
where now ${\mathcal S}_0$ includes both the free action ${\mathcal S}_f$
and quadratic terms responsible of the linear coupling to the mirror:
\begin{eqnarray}\label{eq:defs0}
{\mathcal S}_0(\varphi) &=& 
\frac{1}{2}\int_x (\partial \varphi)^2 
\,+\, \int_{x_\parallel} \Big[ 
\frac{\mu_L}{2} \varphi^2(x_\parallel,0) +  \frac{\mu_R}{2} \varphi^2(x_\parallel,a) \Big] \;,
\end{eqnarray}
where $\mu_{L,R}$ are constants, and ${\mathcal S}_I$ as in (\ref{eq:defsi}).

The known result for ${\mathcal E}_0$, is:
\begin{equation}
	{\mathcal E}_0 \;=\; \frac{1}{2} \int
	\frac{d^3k_\parallel}{(2\pi)^3} \, \log\left[1 \,-\, \frac{e^{- 2
		|k_\parallel| a}}{\big(1 +
			\frac{2|k_\parallel|}{\mu_L}\big) \big(1 +
\frac{2|k_\parallel|}{\mu_R})}\right] \;.
\end{equation}

Again, as in the previous section, we need to evaluate functional averages
of expressions involving powers of the scalar field. Since the functional
weight is again a Gaussian, Wick's theorem for vacuum expectation values
holds true, this time with a different elementary contraction. Therefore,
the evaluation of each term requires the knowledge of $G$, the 2-point
correlation function for the scalar field, in the presence of the Gaussian
weight:
\begin{equation}\label{eq:defcorr}
	G(x;y) \equiv \langle \varphi(x) \varphi(y)\rangle \;,
\end{equation}
where we keep the same symbol for the average as in the previous section,
albeit the functional weight is determined by ${\mathcal S}_0$.

The exact form of $G$ may be explicitly found, and it can be written as follows:
\begin{equation}
	G(x;y) \;=\; G_f(x;y) \,-\, H(x;y) \;,
\end{equation}
where $G_f$ denotes the correlation function corresponding to the free field, i.e., in
the absence of mirrors:
\begin{equation}
	G_f(x;y) \;=\; \langle x| (-\partial^2)^{-1} |y \rangle \;=\; \int
	\frac{d^4k}{(2\pi)^4} \, \frac{e^{i k \cdot (x-y) }}{k^2} \;,
\end{equation}
while
$$
H(x;y) \;=\; \int \frac{d^3k_\parallel}{(2\pi)^3} \frac{dp_3}{2\pi}
	\frac{dq_3}{2\pi} \, e^{i k_\parallel \cdot (x_\parallel -
	y_\parallel)} \frac{1}{(k_\parallel^2 + p_3^2) (k_\parallel^2 +
	q_3^2) D(k_\parallel)} \times
$$
$$
\times \Big\{ \big(\frac{1}{\xi_R(k_\parallel)} +
\frac{1}{2|k_\parallel|}\big) \,
	e^{i (p_3 x_3 - q_3 y_3)} + \big(\frac{1}{\xi_L(k_\parallel)} + \frac{1}{2|k_\parallel|}\big) \, e^{i (p_3 (x_3-a) + q_3 (a - y_3))}
$$
\begin{equation}
- \frac{e^{-|k_\parallel| a}}{2|k_\parallel|} \, 
\big[ e^{i (p_3 x_3 + q_3 (a-y_3))} \,+\, e^{i (p_3 (x_3 - a) - q_3 y_3)} 
\big]  \Big\} \;,
\end{equation}
with
\begin{equation}\label{eq:defd}
	D(k_\parallel) \;=\; 
	\big(\frac{1}{\mu_L} + 
	\frac{1}{2|k_\parallel|}\big) 
	\big(\frac{1}{\mu_R} + \frac{1}{2|k_\parallel|}\big)
	- \frac{e^{- 2 |k_\parallel| a}}{4 k_\parallel^2} \;.
\end{equation}

Note that $G_f(x;y)$ is the limit to which the $G(x;y)$ correlation
function tends when its arguments are at fixed positions with $0 < x_3, y_3
< a$, while the mirrors's positions are infinitely far away. 

There is also another limit that we need to consider when calculating
perturbative corrections: it corresponds to taking the $a \to \infty$
limit, while at the same time assuming that its two arguments belong to one
of the mirrors.  

The results corresponding to the two cases (both arguments on the $L$ or
the $R$ mirror) are, respectively, as follows:
\begin{equation}
	\big[G(x_\parallel,0;y_\parallel,0)\big]_{a \to \infty} \;=\;
\int \frac{d^3k_\parallel}{(2\pi)^3} e^{i k_\parallel \cdot (x_\parallel -
y_\parallel)} \, \frac{1}{2 |k_\parallel| + \mu_L} 
\end{equation}
and
\begin{equation}
	\big[G(x_\parallel,a;y_\parallel,a)\big]_{a \to \infty} \;=\;
\int \frac{d^3k_\parallel}{(2\pi)^3} e^{i k_\parallel \cdot (x_\parallel -
y_\parallel)} \, \frac{1}{2 |k_\parallel| + \mu_R} \;.
\end{equation}

Equipped with the previous ingredients, we now evaluate the explicit form
of the first order terms in the expansion, for the case of quartic
vertices with coefficients $g_L$ and $g_R$.

In the calculation of the first order term, we face the emergence of a
divergence; indeed, we see that:
\begin{equation}\label{eq:first1}
\Gamma_I^{(1)} \;=\; \frac{3}{4!} \, 
\Big\{ 
g_L \int_{x_\parallel} \big[ G(x_\parallel,0;x_\parallel,0) \big]^2
+ 
g_R \int_{x_\parallel} \big[ G(x_\parallel,a;x_\parallel,a) \big]^2
\Big\} \;,
\end{equation}
which, as a rather straightforward calculation shows, is divergent. This
kind of divergence can be dealt with, however, by a similar procedure to
the normal ordering of standard QFT. The important difference here is that
the contraction {\em depends on the distance between the plates\/}, as it is
seen from the form of $G$. 
Therefore, we subtract from the interaction term contributions which
correspond to contractions performed when the planes are infinitely far
apart. In other words, the normal ordering is performed at $a \to \infty$,
so that the interaction actions are renormalized in an intrinsic way for
each mirror,  that is, independently of the presence of the other mirror.

Thus,
\begin{eqnarray}\label{eq:first2}
\Big[\Gamma_I^{(1)}\Big]_{ren} &=& \frac{3}{4!} \, 
g_L \int_{x_\parallel} \Big\{   \big[G(x_\parallel,0;x_\parallel,0) \big]^2 
- 2 \big[G(x_\parallel,0;x_\parallel,0)\big]_{a \to \infty}  
 G(x_\parallel,0;x_\parallel,0)  \nonumber\\
& & + \big[G(x_\parallel,0;x_\parallel,0)\big]_{a \to \infty}^2  
\Big\} \nonumber\\
&+& \frac{3}{4!} \, 
g_R \int_{x_\parallel} \Big\{   \big[G(x_\parallel,a;x_\parallel,a) \big]^2 
- 2 \big[G(x_\parallel,a;x_\parallel,a)\big]_{a \to \infty}  
 G(x_\parallel,a;x_\parallel,a)  \nonumber\\
& & + \big[G(x_\parallel,a;x_\parallel,a)\big]_{a \to \infty}^2  
\Big\} \;.
\end{eqnarray}
The previous expression may also be put in the following form:
\begin{eqnarray}\label{eq:first3}
\Big[\Gamma_I^{(1)}\Big]_{ren} &=& \frac{3}{4!} \, 
g_L \int_{x_\parallel} \big[G_1(x_\parallel,0;x_\parallel,0) \big]^2 \nonumber\\
&+& \frac{3}{4!} \, 
g_R \int_{x_\parallel} \big[G_1(x_\parallel,a;x_\parallel,a) \big]^2 \;,
\end{eqnarray}
with:
\begin{eqnarray}
G_1(x_\parallel,0;x_\parallel,0) &=& G(x_\parallel,0;x_\parallel,0)  
	- \big[ G(x_\parallel,0;x_\parallel,0)\big]_{a \to \infty} \nonumber\\
G_1(x_\parallel,a;x_\parallel,a) &=& G(x_\parallel,a;x_\parallel,a)  
	- \big[ G(x_\parallel,a;x_\parallel,a)\big]_{a \to \infty} \;.
\end{eqnarray}
A rather straightforward calculation shows that:
\begin{eqnarray}
	G_1(x_\parallel,0;x_\parallel,0) &=& \int
	\frac{d^3k_\parallel}{(2\pi)^3} \, \frac{-2 |k_\parallel|}{\mu_L^2} \, 
	\frac{1}{(1 + \frac{2 |k_\parallel|}{\mu_L}) (1 +
		\frac{2 |k_\parallel|}{\mu_R}) e^{ 2 |k_\parallel| a} - 1} \nonumber\\
	G_1(x_\parallel,a;x_\parallel,a) &=& \int
	\frac{d^3k_\parallel}{(2\pi)^3} \, \frac{- 2 |k_\parallel|}{\mu_R^2} \, 
	\frac{1}{(1 + \frac{2 |k_\parallel|}{\mu_L}) (1 +
		\frac{2 |k_\parallel|}{\mu_R}) e^{ 2 |k_\parallel| a} - 1} \;.
\end{eqnarray}

It is worth noting that most of what we have said before could also have
been obtained for the case of {\em momentum dependent\/} coefficients
$\mu_{L,R}$, (with an action which is a straightforward
generalization of the one for constant coefficients).
This allows us to consider a rather `economical' model, consisting of one
where the mass dimensions of the coefficients for the quadratic terms are
given by the momentum itself, so that we are just left with dimensionless
coefficients. Namely, 
\begin{equation}
	\mu_{L,R}(k_\parallel) \;=\; \zeta_{L,R} |k_\parallel|\;,
\end{equation}
where $\zeta_{L,R}$ are dimensionless constants. 

In this case, the first order contribution to the vacuum energy is:
\begin{equation}
	{\mathcal E}^{(1)} = \frac{3}{4! a^4} \, \big[
		\frac{1}{\zeta_L^4} ( 1 + \frac{1}{\zeta_L})^2
		+
		\frac{1}{\zeta_R^4} ( 1 + \frac{1}{\zeta_R})^2
	\big] \, [I(\zeta_L,\zeta_R)]^2 
\end{equation}
where
\begin{equation}
	I(\zeta_L,\zeta_R)\;=\;\frac{1}{4 \pi^2} \int d\rho 
	\rho\,\frac{1}{( 1 + \frac{1}{\zeta_L})( 1 +
	\frac{1}{\zeta_R}) e^{2 \rho} -1} \;. 
\end{equation}
\section{Alternative expansion}\label{sec:strong}
Finally, we will consider here an alternative expansion,
still under the same general structure of the ones considered before,
but such that the interaction terms can (by assumption) be represented in
terms of the generalized Fourier transformations:
\begin{eqnarray}\label{eq:ffourier}
e^{-{\mathcal S}_L(\varphi)} &=& \frac{1}{{\mathcal N}_L} \, \int
{\mathcal D}\xi_L \; e^{- W_L(\xi_L) + i \int_{x_\parallel} \,
\xi_L(x_\parallel) \varphi(x_\parallel,0)} \nonumber\\
e^{-{\mathcal S}_R(\varphi)} &=& \frac{1}{{\mathcal N}_R} \, \int
{\mathcal D}\xi_R \; e^{- W_R(\xi_R) + i \int_{x_\parallel} \,
\xi_R(x_\parallel) \varphi(x_\parallel,a) } \;,
\end{eqnarray}
where $\xi_L$ and $\xi_R$ are auxiliary fields, and ${\mathcal N}_{L,R}$
are normalization constants.

It may be thought of as a particular case of the previous section, namely,
when the constants $\mu_L$ and $\mu_R$ tend to infinity,  while the 
microscopic interactions are such that, there still are non-vanishing
small non-quadratic functions $W_{L,R}$.

We then insert the representations (\ref{eq:ffourier}) into the definition
(\ref{eq:defz}) of ${\mathcal Z}$, and integrate out the scalar field
$\varphi$, to obtain:
\begin{equation}
{\mathcal Z} \;=\; \frac{{\mathcal Z}_f}{{\mathcal N}_L{\mathcal N}_R} \,
\int {\mathcal D}\xi_L {\mathcal D}\xi_R \, e^{-W_q(\xi_L,\xi_R) - W_L(\xi_L) - W_R(\xi_R)} 
\end{equation}
where $W_q(\xi_L,\xi_R)$ denotes the quadratic form in the auxiliary
fields: 
\begin{equation}
W_q(\xi_L,\xi_R)\;=\; \frac{1}{2} \, \int_{x_\parallel,x'_\parallel} \,
\xi_A(x_\parallel) \, K_{AB}(x_\parallel,x'_\parallel) \,
\xi_B(x'_\parallel) \;\;,\;\; A = L,\, R\;, 
\end{equation}
where 
\begin{equation}
K_{AB}(x_\parallel,x'_\parallel) \;=\; \int
\frac{d^3k_\parallel}{(2\pi)^3} \, e^{i k_\parallel \cdot
(x_\parallel-x'_\parallel)}\, {\widetilde K}_{AB}(k_\parallel)
\end{equation}
with:
\begin{equation}
\big[{\widetilde K}_{AB}(k_\parallel)\big] \;=\; \frac{1}{2 |k_\parallel|} \, 
\left(
\begin{array}{cc}
1 & e^{-|k_\parallel| a} \\
e^{-|k_\parallel| a} & 1
\end{array} 
\right) \;.
\end{equation}
Proceeding in an analogous manner to the one followed for the weak
coupling expansion, but with $W_q$ playing a role similar to ${\mathcal
S}_f$, we see that:
\begin{equation}
\Gamma \;=\; \Gamma_q \,+\, \Gamma_s 
\end{equation}
with 
\begin{equation}
e^{-\Gamma_q} \;=\; \int {\mathcal D}\xi_L {\mathcal D}\xi_R \, e^{-W_q(\xi_L,\xi_R)} 
\end{equation}
and
\begin{eqnarray}
e^{-\Gamma_s} &=& \langle e^{- W_L(\xi_L) - W_R(\xi_R)} \rangle_q
\nonumber\\
\langle \ldots \rangle_q &\equiv& \frac{\int {\mathcal D}\xi_L {\mathcal
D}\xi_R \ldots e^{-W_q(\xi_L,\xi_R)}}{\int {\mathcal D}\xi_L {\mathcal
D}\xi_R e^{-W_q(\xi_L,\xi_R)}} \;.
\end{eqnarray}
It is worth noting that, as well as for the weak coupling case, there will
be self-energies coming from the subleading terms in the
series. A convenient way to get rid of them is to apply a normal ordering,
in such a way that $a$ independent contributions coming from contractions
between points on the same vertex are discarded. In the weak coupling case,
that amounts to the usual normal ordering, since the free propagator
is independent of $a$. Here, on the contrary, the propagator for the
auxiliary fields does depend on $a$. Thus, the normal ordering subtracts
from it a contribution which is independent of $a$. More explicitly, when
including tadpoles, they contribute:
\begin{equation}
\langle \xi_L(x_\parallel) \xi_L(x_\parallel) \rangle_q \;=\;
\int \frac{d^3k_\parallel}{(2\pi)^3} \frac{2 |k_\parallel|}{e^{2
|k_\parallel|a}-1}\,=\, \langle \xi_R(x_\parallel) \xi_R(x_\parallel) \rangle_q
\;.
\end{equation}

The leading term is then $\Gamma_q$ which is identical
to the Dirichlet result, and therefore the corresponding energy per unit
area is given by:
\begin{equation}
{\mathcal E}_q\;=\; -\frac{\pi^2}{1440 \, a^3} \;.
\end{equation}

Let us now consider the expansion of $\Gamma_s$,  which contains the
subleading terms. Note that, since now the
Gaussian average depends on $a$, even the first order terms in $W_{L,R}$ may produce
non-trivial contributions to the interaction energy. Indeed, we see that:
\begin{equation}
\Gamma_s^{(1)}\;=\;\langle W_L(\xi_L) \rangle_q \,+\,\langle W_R(\xi_R)
\rangle_q \;.
\end{equation}

As an example, let us consider:
\begin{equation}
 W_R(\xi_R) \;=\; \int_{x_\parallel} \, \frac{g_R}{4!}\,
[\xi_R(x_\parallel)]^4 \;,  W_L(\xi_L) \;=\; 0\;,
\end{equation}
namely, a nonlinear mirror at $x_3=a$ and a Dirichlet one at $x_3=0$. In this case, we obtain,
\begin{eqnarray}
\Gamma_s^{(1)}&=& \frac{g_R}{8} \,
\left(\int \frac{d^3k_\parallel}{(2\pi)^3} \frac{2 |k_\parallel|}{e^{2
|k_\parallel|a}-1} \right)^2 \nonumber\\
&=& \frac{\pi^6}{460800 g_R \, a^8} \;.
\end{eqnarray}

\section{Conclusions}\label{sec:conclusions}
We have presented a perturbative treatment for the calculation of the Casimir
energy in a system where the boundary conditions, imposed on a real scalar
field, are nonlinear. This has the consequence that, for example, there is
no Lifshitz formula~\cite{lifshitz} to account for the Casimir interaction
energy, since the mirrors' properties cannot be represented just by
reflection and transmission coefficients when they are nonlinear.  

We have considered different situations: first the case of semitransparent
nonlinear mirrors, then mirrors including both linear and nonlinear
contributions in their boundary conditions, and finally situations where
the mirrors can be described as having a small nonlinear contribution on
top of an otherwise perfect (Dirichlet) plane.

The nonlinearities manifest themselves, at this level, in the presence of
terms in the interaction energy which have a non standard dependence on the
distance $a$, even when they are semitransparent.

\section*{Acknowledgements}
C.D.F has been supported by CONICET, ANPCyT and UNCuyo, Argentina.
The Conselho Nacional de Desenvolvimento Científico e Tecnológico
(CNPq-Brazil) is also acknowledged for financial support.



\begin{thebibliography}{bib}
\bibitem{rev}
G. Plunien, B. M\"uller, and W. Greiner, Phys. Rep. \textbf{134},
87 (1986); P. Milonni, {\it The Quantum Vacuum} (Academic Press,
San Diego, 1994); V. M. Mostepanenko and N. N. Trunov, {\it The
Casimir Effect and its Applications} (Clarendon, London, 1997); M.
Bordag, {\it The Casimir Effect 50 Years Later} (World Scientific, Singapore, 1999);
M. Bordag, U. Mohideen, and V. M. Mostepanenko, Phys. Rep.
\textbf{353}, 1 (2001); K. A. Milton, {\it The Casimir Effect:
Physical Manifestations of the Zero-Point Energy} (World
Scientific, Singapore, 2001); S. Reynaud {\it et al.}, C. R. Acad.
Sci. Paris \textbf{IV-2}, 1287 (2001); K. A. Milton, J. Phys. A:
Math. Gen. \textbf{37}, R209 (2004); S.K. Lamoreaux, Rep. Prog.
Phys. \textbf{68}, 201 (2005); Special Issue {\it "Focus on Casimir Forces"},
New J. Phys. \textbf {8} (2006).
\bibitem{Elizalde:1995bb}
E.~Elizalde and S.~D.~Odintsov,
Class.\ Quant.\ Grav.\  {\bf 12}, 2881 (1995)\;.
\bibitem{Saharian:2006zm}
A.~Saharian and G.~Esposito,
J.\ Phys.\ A  {\bf 39}, 5233 (2006);\\
 ibidem, 
{\it In the Proceedings of 11th Marcel Grossmann Meeting on General Relativity, Berlin, Germany, 23-29 Jul 2006, pp 2761-2763}.
\bibitem{Fosco:2009cw} 
C.~D.~Fosco and E.~Losada,
Phys.\ Lett.\ B {\bf 675}, 252 (2009).
\bibitem{EsquivelSirvent:2006iw}
 R.~Esquivel-Sirvent, C.~Villarreal, W.~L.~Moch\'an, A.~M.~Contreras-Reyes
and V.~B.~Svetovoy, J.\ Phys.\ A  
{\bf 39} (2006) 6323;\\
 R.~Esquivel-Sirvent, C.~Villarreal and W.~L.~Moch\'an, Phys.\ Rev.\ A
{\bf 68} 052103.\\
 R.~Esquivel-Sirvent, C.~Villarreal and W.~L.~Moch\'an, Phys.\ Rev.\ A {\bf
71} 029904;\\
 R.~Esquivel-Sirvent and W.~L.~Moch\'an, {\em Quantum Field Theory Under
the Influence of External Conditions}, Ed.~ K.~Milton (New Jersey: Rinton).
\bibitem{bajnok}Z.~Bajnok, L.~Palla and G.~Tak\'acs, Phys.\ Rev.\ {\bf D73} 065001 (2006), \\
Z.~Bajnok, L.~Palla and G.~Tak\'acs, Nucl.\ Phys.\ {\bf B772} 290 (2007).
\bibitem{Graham:2003ib}
  See, for example:\\
N.~Graham, R.~L.~Jaffe, V.~Khemani, M.~Quandt, O.~Schroeder and H.~Weigel,
  Nucl.\ Phys.\  B {\bf 677}, 379 (2004), 
and references therein.
\bibitem{Jaffe:2003ji}
R.~L.~Jaffe,
AIP Conf.\ Proc.\  {\bf 687}, 3 (2003)\;.
\bibitem{Fosco:2008td}
  C.~D.~Fosco, F.~C.~Lombardo and F.~D.~Mazzitelli,
  Phys.\ Lett.\  B {\bf 669}, 371 (2008)
  [arXiv:0807.3539 [hep-th]].
\bibitem{Fosco:2009zc} C.~D.~Fosco, F.~C.~Lombardo and F.~D.~Mazzitelli,
Phys.\ Rev.\ D {\bf 80}, 085004 (2009).
\bibitem{Milton:2008vr} K.~A.~Milton, P.~Parashar and J.~Wagner,
Phys.\ Rev.\ Lett.\  {\bf 101}, 160402 (2008).
\bibitem{Wagner:2008qq} J.~Wagner, K.~A.~Milton and P.~Parashar,
J.\ Phys.\ Conf.\ Ser.\  {\bf 161}, 012022 (2009).
\bibitem{ZinnJustin:2002ru} 
J.~Zinn-Justin,
Int.\ Ser.\ Monogr.\ Phys.\  {\bf 113}, 1 (2002).
\bibitem{lifshitz} E.M. Lifshitz, Zh. Eksp. Teor. Fiz. {\bf 29}, 94 (1955) [Sov. Phys. JETP
{\bf 2}, 73 (1956).
\end{thebibliography}
\end{document}